\documentclass[twocolumn,preprintnumbers,superscriptaddress,nofootinbib,showpacs,showkeys]{revtex4}
\usepackage{amsmath}
\usepackage{amssymb}
\usepackage{graphicx}
\usepackage{hyperref}
\usepackage{xspace}
\usepackage{bm}
\usepackage{slashed}
\usepackage{dsfont}
\usepackage{bbm}

\usepackage{booktabs}
\AtBeginDocument{
\heavyrulewidth=.08em
\lightrulewidth=.05em
\cmidrulewidth=.03em
\belowrulesep=.65ex
\belowbottomsep=0pt
\aboverulesep=.4ex
\abovetopsep=0pt
\cmidrulesep=\doublerulesep
\cmidrulekern=.5em
\defaultaddspace=.5em
}

\newcommand{\be}{\begin{equation}}
\newcommand{\ee}{\end{equation}}
\newcommand{\bea}{\begin{eqnarray}}
\newcommand{\eea}{\end{eqnarray}}

\usepackage{xcolor}
\definecolor{light-gray}{gray}{0.8}

\begin{document}

\title{From Pentaquarks to Dibaryons in ${\bm \Lambda_b(5620)}$ decays}


\newcommand{\CERNaff}{CERN, PH-TH, CH-1211 Geneva 23, Switzerland}
\newcommand{\sapienza}{Dipartimento di Fisica and INFN, `Sapienza' Universit\`a di Roma\\
P.le Aldo Moro 5, I-00185 Roma, Italy}
\newcommand{\columbia}{Department of Physics, 538W 120th Street,
Columbia University, New York, NY, 10027, USA}
\newcommand{\pavia}{INFN Pavia, Via A. Bassi 6, I-27100 Pavia, Italy}
\newcommand{\romadue}{Dipartimento di Fisica and INFN, Universit\`a di Roma `Tor Vergata'\\
 Via della Ricerca Scientifica 1, I-00133 Roma, Italy}
\newcommand{\alice}{ALICE\xspace}

\author{L. Maiani }
\affiliation{\sapienza}
\author{A.D.~Polosa}
\affiliation{\sapienza}
\author{V.~Riquer}
\affiliation{\sapienza}

\begin{abstract}
Pentaquarks and dibaryons are natural possibilities if diquarks are used as the building blocks to assemble hadrons. In this short note, motivated by the very recent discovery of two pentaquark states, we highlight some possible channels to search for dibaryons in $\Lambda_b(5620)$ decays. 
 \end{abstract}


\maketitle
Color antisymmetric diquarks can replace anti-quarks in conventional hadrons, to give conventional, but also unconventional new hadrons. 
Indeed, as for the color charge $\alpha$, a quark $q^\alpha$ is equivalent to an antidiquark $\bar{\bm d}^\alpha=[\bar q\bar q^\prime]_{\bm 3}^\alpha$, and an antiquark $\bar q_\alpha$ to a diquark $\bm d_\alpha=[qq^\prime]_{{\bar{\bm 3}},\,\alpha}$, as long as attraction in the $\bm 3(\bar{\bm 3})$ channels is assumed. 

Starting from an antibaryon, we get in first step: $\bar q \bar q^\prime [q_1 q_2]$, namely a tetraquark, $X, Y, Z$. In the next step we obtain the pentaquark $\bar q [q_1 q_2][q_3 q_4]$.

The final step is one in which we replace all antiquarks in the baryon, saturating completely with diquarks the junction of three colored strings represented by the antisymmetric color tensor, $\epsilon^{\alpha \beta \gamma}$. We obtain in this way a dibaryon, the $B=2$ color bound alternative to the deuteron, with all its strange, charmed etc. variations. It seems a reasonable possibility that tetraquarks, pentaquarks and dibaryons make the next layer of hadron spectroscopy following the first layer made by the Gell-Mann--Zweig baryons and mesons.

Dibaryons were envisaged by Jaffe~\cite{Jaffe} to bind 6 quarks in a stable $0^+$ flavor singlet at a mass of about 2000~MeV (called a H-dihyperon, later dibaryon). An early discussion based on the Dolen-Horn-Schmidt duality between baryon-(anti)baryon scattering and annihilation channels~\cite{DHS} is found in~\cite{Veneziano} and `baryonia' are discussed in~\cite{Montanet}.  For a recent lattice QCD study of baryon-baryon interactions see~\cite{HAL}.

Dibaryons at about~2~GeV have been considered in a number of papers, usually as 6-quark states in a MIT bag, 
see~\cite{Mulders,Balachandran}. 
Diquarks have been used by Jaffe and Wilczek~\cite{jw} to describe complex hadron structures like the (later disproved) `old pentaquark'. Hidden charm ``hexaquarks'' are also discussed in~\cite{brleb}.

Along similar lines, heavy-light diquarks were introduced in~\cite{noi1} with interesting consequences for the description of the $X(3872)$ meson and related $X$, $Y$ and $Z$ states as tetraquarks, with hidden charm or beauty~\cite{tipo2,ali}. For a review see~\cite{revs}.
The recently discovered heavy pentaquarks~\cite{pentaquarks} (with masses $\sim 4400$~MeV) are explained within a natural extension of this scheme~\cite{noi2}, see also~\cite{altr}.

The lightest charmed dibaryon may be observed in $\Lambda_b$ decay, already a source of pentaquarks.

We start with the Cabibbo allowed decay, adding two light pairs from the vacuum
\be
\Lambda_b(bud) \to c d \bar u +ud+(u\bar u d \bar d)_{\rm vac} \nonumber
\ee
that gives
\bea
&&\Lambda_b \to \bar p + [cd][ud][ud]=\bar p +{\mathds{D}_c^+}\notag\\
&& M({\mathds{D}_c^+})< 4682~{\rm MeV}
\label{uno}
\eea 

The decay of the charmed dibaryon, ${\mathds{D}_c^+}$, may take different routes, according to its mass, in relation to pentaquark masses. The preferred decay would be by string breaking, into a baryon plus a pentaquark.  However it is  possible that this route is forbidden by energy conservation, even for the lightest, spin $1/2$ pentaquarks~\cite{noi2}. Indeed, the known $X,Y,Z$, with the exception of $Y(4630)$~\cite{cot}, do not decay into baryon-antibaryon pairs (string breaking) but rather into charmonium plus meson (quark rearrangement). Similarly, the observed pentaquarks  do not decay into the channels preferred by string breaking, such as $X(Y)$ plus proton, forbidden by energy conservation, but in the quark rearrangement channel, $J/\Psi+p$.

We analyze in sequence the possible decay chains. 

{\bf \emph{Decay by quark rearrangement.}} 
For analogy with the observed tetraquark and pentaquark decay, we put in the first line the quark rearrangement decays
\bea
&& {\mathds{D}_c^+} =[cd][ud][ud] \to \nonumber \\
&& \to p + \Sigma_c^0 (\to p+ \Lambda_c ^+ +\pi^-)~~~{\rm or}~~~n+\Lambda_c^+ 
\label{qrearr}
\eea 
Note the occurrence of $\Sigma_c^0$ in the first decay, necessary if a proton is required for the lack of visibility of the neutron. 

{\bf \emph{Decay by string breaking.}}
Breaking one color string by a $u \bar u$ pair, a possible decay path is
\be
{\mathds{D}_c^+}\to p+ \mathds{P}_c^0(\bar u [cd][ud]) 
\label{due}
\ee
with the final charmed pentaquark decaying as
\be
{\mathds{P}_c^0}\to  \Lambda_c^+ + \pi^-~~~{\rm or}~~~{\mathds{P}_c^0}\to n+D^0
\label{tre}
\ee

Another experimental signature is obtained with a $s\bar s$ pair from the vacuum, replacing step (\ref{due}) by
\be
{\mathds{D}_c^+}\to \Lambda+{\mathds{P}_{c\bar s}^+}(\bar s [cd][ud])
\label{duestra}
\ee
followed by
\be
{\mathds{P}_{c\bar s}^+}\to K^0 + \Lambda_c^+
\ee

{\bf \emph{Overall $\Lambda_b$ decay chains.}}
Discarding decay channels with a neutron, the interesting $\Lambda_b$ decay chain in (\ref{qrearr}) and (\ref{tre}) is
\be
\Lambda_b \to  \bar p + p+\Lambda_c^+ + \pi^-,~~~M({\mathds{D}_c^+})> 3364~{\rm MeV} \label{lambdac}
\ee
with (\ref{qrearr}) and (\ref{tre}) distinguished by the occurrence of a pentaquark resonance or of the $\Sigma_c^0$ in the $\Lambda_c^+ \pi^-$ channel. 

The case (\ref{duestra}) leads to
\bea
&&\Lambda_b \to \bar p + \Lambda+\Lambda_c^+ +K^0\nonumber \\
&&M({\mathds{D}_c^+})>M(\Lambda)+M({\mathds{P}_{c\bar s}^+})>3901~{\rm MeV}\label{kappazero}
\eea

{\bf \emph{Weak semileptonic decays.}}
For dibaryon mass below the limit in (\ref{lambdac}), $\beta$-decay of the charm quark allows the dibaryon to transform into uncharmed baryon pairs with strangeness $S=-1,0$ according to

\paragraph*{{\rm Cabibbo~ allowed decay:}}
\bea
&&{\mathds{D}_c^+}\to e^+ +\nu_e +\Sigma^- + p\nonumber \\
&& M({\mathds{D}_c^+})>2136~{\rm MeV} \label{caballowd}
\eea
\paragraph*{{\rm Cabibbo~ forbidden decays:}}
\bea
&&{\mathds{D}_c^+}\to e^+ +\nu_e +\Delta^- + p\nonumber \\
&& M({\mathds{D}_c^+})>2174~{\rm MeV}
\label{cabfbd2}
\eea

\bea
&&{\mathds{D}_c^+}\to e^+ + \nu_e + 2n\nonumber \\
&& M({\mathds{D}_c^+})> 1879~{\rm MeV}
\label{cabfbd3}
\eea
For masses below $1879$~MeV the lightest charmed dibaryon is stable.

If diquarks are good building blocks to assemble hadrons, the missing structures to discover, after pentaquarks and tetraquarks,  are dibaryons. 
At the LHCb, the lightest charmed dibaryon maybe searched in $\Lambda_b$ decays: $\bm{ (i)}$ in the mass range $4682~{\rm MeV}> M({\mathds{D}_c^+}) > 3364~{\rm MeV}$, among the particles recoling against an antiproton as indicated in (\ref{lambdac}), or, $\bm{(ii)}$ in the range $3364~{\rm MeV}>M({\mathds{D}_c^+})>1879~{\rm MeV}$, searching for weakly decaying particles, see Eqs. (\ref{caballowd},\ref{cabfbd2},\ref{cabfbd3}), with the typical lifetime of charmed particles.

We are aware that tetraquark and pentaquark states identified until now all involve heavy quark flavours and that systematically replacing $q^\alpha \to {\bm {\bar d}}^\alpha$ in light systems will give rise to exotic mesons and baryons never clearly identified. If this is due to experimental reasons, such as too large widths, or to peculiarities of the strong interaction forbidding the binding of the light `bad diquarks' under certain circumstances, will need a closer examination.  

We acknowledge private communication on channel visibility in LHCb with S.~Stone and T.~Skwarnicki.

\end{document}